\documentclass[lettersize,journal]{IEEEtran}
\usepackage{algorithm}
\usepackage{algorithmicx}
\usepackage{algpseudocode}
\usepackage{array}
\usepackage{amsmath}
\usepackage{amssymb}
\usepackage[caption=false,font=normalsize,labelfont=sf,textfont=sf]{subfig}
\usepackage{bm}
\usepackage{booktabs}
\usepackage{color}
\usepackage{textcomp}
\usepackage{stfloats}
\usepackage{url}
\usepackage{cite}
\usepackage{multirow}
\usepackage[colorlinks,hyperindex,breaklinks]{hyperref}
\usepackage{pifont}
\usepackage{graphicx}
\usepackage{ragged2e}
\usepackage{arydshln}

\newcommand{\etal}{~\textit{et al}.}
\newcommand{\ie}{\textit{i}.\textit{e}.}
\newcommand{\eg}{\textit{e}.\textit{g}.}
\begin{document}
\title{LENAS: Learning-based Neural Architecture Search and Ensemble for 3D Radiotherapy Dose Prediction}

\author{Yi Lin, Yanfei Liu, Hao Chen, Xin Yang, Kai Ma, Yefeng Zheng, Kwang-Ting Cheng
\thanks{This work was supported by the Shenzhen Science and Technology Innovation Committee Fund (Project No. SGDX20210823103201011) and Hong Kong Innovation and Technology Fund (Project No. ITS/028/21FP).}

\thanks{Y. Lin, H. Chen, and K.-T. Cheng are with the Department of Computer Science and Engineering, The Hong Kong University of Science and Technology, Hong Kong. H. Chen is also with the Department of Chemical and Biological Engineering, The Hong Kong University of Science and Technology. (e-mails: yi.lin@connect.ust.hk; jhc@cse.ust.hk; timcheng@ust.hk).}

\thanks{Y. Liu is with Shenzhen United Imaging Research Institute of Innovative Medical Equipment, Real-Time Laboratory, Shenzhen 518048, China. (e-mail: yanfei.liu@cri-united-imaging.com).  }

\thanks{X. Yang is with the Department of Electronic Information and Communications, Huazhong University of Science and Technology, Wuhan 430074, China. (e-mail: xinyang2014@hust.edu.cn).  }

\thanks{K. Ma, and Y. Zheng are with Jarvis Research Center, Tencent YouTu Lab, Shenzhen 518057, China. (e-mails: \{kylekma, yefengzheng\}@tencent.com).}

\thanks{The first two authors contributed equally and H. Chen is the corresponding author.}}

\markboth{Accepted by IEEE Transactions on Cybernetics}%
{Lin \MakeLowercase{\textit{et al.}}: LENAS}
\maketitle
\begin{abstract}
Radiation therapy treatment planning requires balancing the delivery of the target dose while sparing normal tissues, making it a complex process.
To streamline the planning process and enhance its quality, there is a growing demand for knowledge-based planning (KBP).
Ensemble learning has shown impressive power in various deep learning tasks, and it has great potential to improve the performance of KBP.
However, the effectiveness of ensemble learning heavily depends on the diversity and individual accuracy of the base learners.
Moreover, the complexity of model ensembles is a major concern, as it requires 
maintaining multiple models during inference, leading to increased computational cost and storage overhead.
In this study, we propose a novel learning-based ensemble approach named LENAS, which integrates neural architecture search with knowledge distillation for 3D radiotherapy dose prediction. 
Our approach starts by exhaustively searching each block from an enormous architecture space to identify multiple architectures that exhibit promising performance and significant diversity.
To mitigate the complexity introduced by the model ensemble, we adopt the teacher-student paradigm, leveraging the diverse outputs from multiple learned networks as supervisory signals to guide the training of the student network.
Furthermore, to preserve high-level semantic information, we design a hybrid-loss to optimize the student network, enabling it to recover the knowledge embedded within the teacher networks.
The proposed method has been evaluated on two public datasets, OpenKBP and AIMIS. Extensive experimental results demonstrate the effectiveness of our method and its superior performance to the state-of-the-art methods. 
Code: \url{github.com/hust-linyi/LENAS}.
\end{abstract}
\section{Introduction}
\label{sec:introduction}
\IEEEPARstart{R}{adiation} therapy, chemotherapy, surgery, or their combination are widely utilized in clinical settings for cancer control~\cite{park2019imaging}. 
Compared to conventional 3D conformal therapy, modern treatment methods such as intensity modulation radiation therapy and volumetric arc therapy place greater emphasis on delivering the prescribed dosage to the planning target volume (PTV) while minimizing radiation exposure to organs-at-risk (OAR)~\cite{Rifat}. 
Given the proximity of tumors to these critical structures, accurate tumor delineation is crucial to prevent radiation-induced toxicity~\cite{lin2019deep}.
The process of achieving a treatment plan with an optimal dose distribution involves meticulous iterations by a physicist, who adjusts various treatment planning parameters and weightings to balance clinical objectives~\cite{nguyen20193d}. 
However, this procedure is time-consuming and prone to inter- and intra-observer variability, given the varying experience and skills of physicists~\cite{ye2017collision}.

Knowledge-based planning (KBP) presents an automatic solution that addresses the human effort required in traditional treatment planning by generating dose distributions, patient-specific dose-volume histograms (DVH), and dose constraints for PTVs and OARs~\cite{zhou2023all,yu2023first,fan2019automatic}. 
This approach serves as a reference to optimize the planning and control its quality, which effectively streamlines the treatment planning process.
Recently, the field of KBP has witnessed significant advancements inspired by deep learning techniques~\cite{huynh2020artificial}. 
Researchers have explored data-driven approaches to directly predict dose distributions. For instance, Nguyen\etal~\cite{nguyen2019feasibility} utilized U-Net to predict dose distributions for prostate cancer, while Fan\etal~\cite{fan2019automatic} extended this approach by employing ResUNet for dose prediction in head-and-neck cancer. Kandalan\etal~\cite{kandalan2020dose} investigated the generalizability of U-Net for prostate cancer dose prediction through transfer learning with limited input data.
However, existing methods typically rely on U-Net and its variants~\cite{nguyen20193d,nguyen2019feasibility}, which may not guarantee applicability across different physicists, diseases, and clinical settings. 

Ensemble learning has been widely adopted in various deep learning tasks and has demonstrated impressive power~\cite{zolfaghari2023cancer}.
These ensembles consist of multiple neural networks, with predictions combined through weighted averaging or voting during the test stage. 
While diversity is considered crucial for ensemble learning~\cite{bian2021does}, existing methods often overlook it by employing a single network architecture coupled with various training strategies or the combination of off-the-shelf architectures.
The issue of limited diversity can be effectively addressed by leveraging neural architecture search (NAS) method, which can generate a large number of diverse architectures, driving a natural bias towards diversity of predictions, and in turn to afford the opportunity to integrate these networks to achieve imporved results. 
However, several important research gaps are rarely explored:
First, the relative importance of base learners' performance and diversity in ensemble learning is not well understood.
Second, striking a balance between ensemble performance and computational complexity is another crucial aspect that demoands further investigation.
Third, encouraging diversity within the search process of NAS deserves attention.
\begin{figure*}[t!]
    \centering
    \includegraphics[width=0.95\textwidth]{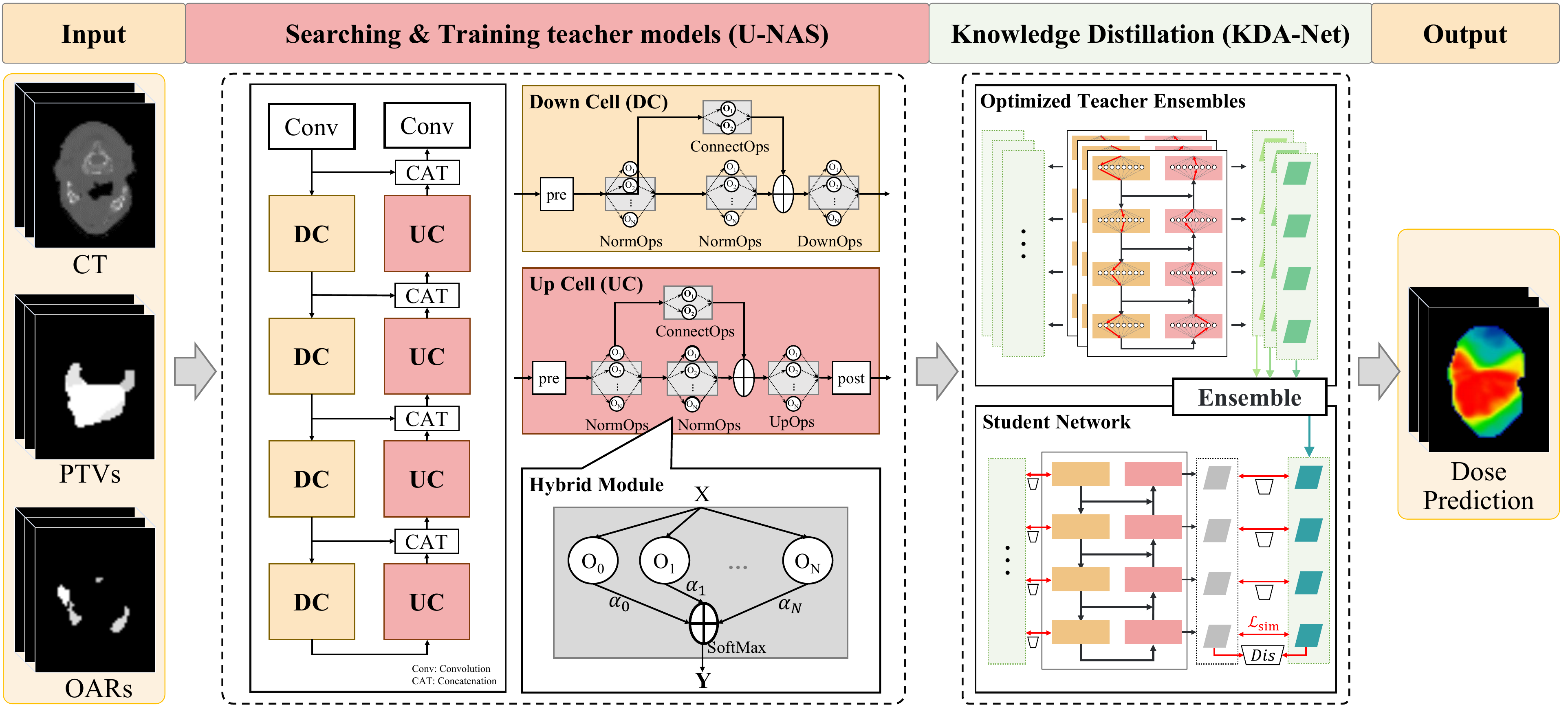}
    \caption{Overview of the proposed LENAS.
    $O_i$ in the hybrid module denotes the operation and $\alpha_i$ denotes its weight. $Dis$ in KDA-Net denotes the discriminator.}
    \label{fig_framwork}
\end{figure*}
To address the aforementioned challenges, we propose a learning-based ensemble approach with NAS for dose prediction, named LENAS. 
Our method adopts the teacher-student paradigm by leveraging a combination of diverse outputs from multiple automatically designed neural networks as a teacher model zoo to guide the target student network. 
The core of our LENAS includes two modules. 
First, instead of relying on off-the-shelf networks, we present a novel U-shape differentiable neural architecture search framework, named U-NAS, which automatically and efficiently searches for neural architectures from enormous architecture configurations for the teacher ensembles. 
In addition, we design a diversity--encouraging loss to ensure both the high performance and diversity of the searched architectures.
Second, to reduce the computational costs in the inference phase and meanwhile ensure high ensemble performance, we further present a knowledge distillation (KD) network with adversarial learning, named KDA-Net, which hierarchically transfers the distilled knowledge from the teacher networks to the student network.

The proposed methodology is evaluated using two publicly available datasets, the OpenKBP dataset from the 2020 AAPM Grand Challenge and the AIMIS dataset from the 2021 Tencent AIMIS Challenge (Task 4). 
Our U-NAS framework achieves exceptional performance on the OpenKBP dataset, surpassing the performance of the champion of the AAPM challenge~\cite{babier2020openkbp,long2015fully,milletari2016v,ronneberger2015u}.
Furthermore, our single LENAS model also outperformed state-of-the-art methods, securing first place in the AIMIS challenge.
Our contributions are five-fold:
\begin{itemize}
    \item We introduce LENAS, a learning-based ensemble framework, which comprises the U-NAS framework for efficient and automatic architecture search, and KDA-Net for achieving a balance between efficiency and accuracy.
    \item We propose a diversity--encouraging loss, which explicitly enhances the diversity among the searched models.
    \item We design a hybrid--loss that facilitates hierarchical transfer of knowledge from teacher ensembles to a single lightweight model, containing both hard (similarity loss) and soft (adversarial loss) constraints.
    \item We provide in-depth analyses and empirical guidelines for generating and selecting base learners in ensemble.
    \item Extensive experiments on two public datasets demonstrated the effectiveness of each module and the superior performance of our method over state-of-the-art methods. 
\end{itemize}
\section{Related Work}
\label{sec:2}
\subsection{Knowledge-Based Planning}
\label{sec:2.1}
Knowledge-based planning (KBP) is realized by building an atlas-based repository or a mathematical model to predict the dosimetry such as dose distribution, entire DVH curve, and dose volume metrics utilizing previously optimized plans~\cite{momin2021knowledge}. 
In atlas-based methods, manually designed geometric features are selected as metrics to measure the similarity between previous plans and a new plan. 
The previous parameters of the most similar plan are adopted as the initialization of the new plan optimization. 
Conversely, modeling methods use handcrafted features to regress and predict DVH of a new plan to guide the optimization processing~\cite{zhu2011planning}. 
The features include overlap volume histogram (OVH)~\cite{li2022explainable}, beams eye view (BEV) projections, and overlap of regions of interest (ROIs), etc., which are applicable to both atlas-based and modeling methods.

Nevertheless, traditional knowledge-based planning (KBP) methods have limitations as they only predict 2-dimensional or 1-dimensional dosimetry metrics, failing to capture the complete spatial distribution of dosage.
In recent years, researchers mainly focused on deep learning-based KBP methods. 
Leveraging the powerful capabilities of CNNs to extract statistical and contextual features, these methods enable direct and highly accurate prediction of 3-dimensional voxel-wise dose distributions. 
The inputs of deep learning-based models usually are images (\eg, CT images and structure masks), and the architecture of models is mainly U-Net~\cite{nguyen20193d}. 
The two main directions for improving the performance of CNN-based dose prediction are: 1) designing different architectures, including modified U-Net~\cite{lin2023rethinking}, U-Res-Net~\cite{liu2019deep}, HD U-Net~\cite{nguyen20193d}, GAN-based~\cite{nguyen2020incorporating}; 
2) adding clinical parameters into inputs, such as isocenter~\cite{willems2019feasibility}, beam geometry infomation~\cite{teng2024beam}, isodose lines and gradient infomation~\cite{tan2021incorporating}).

\subsection{Ensemble Learning}
Ensemble learning has shown impressive power in various deep learning tasks~\cite{zolfaghari2023cancer}, a large amount of literature has provided theoretical and empirical justiﬁcations for its success, including Bayesian model averaging~\cite{yang2023survey}, enriching representations~\cite{yang2021two}, and reducing stochastic optimization error~\cite{zhou2021ensemble}. 
These arguments reached a consensus that the individual learner in the ensembles should be \textit{accurate and diverse}~\cite{bian2021does,wood2023unified}. 
To encourage the diversity of the ensembles, the strategies for building ensembles typically include: 1) training the same network with various settings, such as bagging~\cite{altman2017ensemble}, random initializations~\cite{lin2023nuclei}, and different hyper-parameters~\cite{mohammed2023comprehensive} (\eg, iteration, learning rate, and objective function); and 2) training different networks with the various architectures~\cite{lin2019seg4reg}. One of the most famous techniques is dropout~\cite{srivastava2014dropout}, in which some of the neurons are dropped in each iteration, and the final model can be viewed as an ensemble composed of multiple different sub-models. 

For combining the predictions of each base model in an ensemble, the most prevalent method is majority voting~\cite{campagner2023aggregation} for classification and segmentation (which can be viewed as pixel-wise classification), and simple averaging~\cite{brown2005managing} for the regression task. 
Despite their success, most existing ensemble methods do not explicitly balance the two important factors, \ie, the accuracy of individual learners and diversity among them. 
Different from these attempts, we reveal that the ensemble candidates produced by NAS could simultaneously guarantee the diversity and individual model's accuracy, achieving superior ensemble performance.

\subsection{Neural Architecture Search}
Neural architecture search (NAS) aims at searching for a desirable neural architecture from a large architecture collection.
It has received increasing interest in various medical image analysis tasks, such as image classification~\cite{dondeti2020deep}, localization~\cite{jiang2020elixirnet}, segmentation~\cite{weng2019unet}, and reconstruction~\cite{yan2020neural}. 
Much of the focus has been on the design of search space and search strategy. 
For example, Weng\etal\cite{weng2019unet} introduced NAS-Unet for 2D medical image segmentation, which consists of different primitive operation sets for down- and up-sampling cells.
Zhu\etal\cite{zhu2019v} proposed V-NAS for volumetric medical image segmentation that built a search space including 2D, 3D, or pseudo-3D (P3D) convolutions. 
As for the search strategy, existing research can be categorized into three classes: the evolutionary algorithm~\cite{real2019regularized}, reinforcement learning~\cite{zoph2018learning}, and gradient-based differentiable methods~\cite{liu2019darts}. 

In addition to searching for the single best model, recent works~\cite{zaidi2021neural,bian2021subarchitecture,shu2021going} proposed to search diverse architectures to form stronger and more robust ensembles. 
For instance, NEAS~\cite{chen2021one} and SAEP~\cite{bian2021subarchitecture} encourage the disagreement between the base classifiers for ensemble pruning. NES~\cite{zaidi2021neural} and NESS~\cite{shu2021going} optimize the sampled sub-architecture ensembles based on the evolutionary or sampling algorithm. 
Our method is different from the methods above in three ways. 
First, the ensemble pruning methods~\cite{bian2021subarchitecture,hu2020angle} attempt to reduce the search space to facilitate the search process, while our method directly distills large ensembles into a single network. 
Second, the neural ensemble search methods~\cite{zaidi2021neural,shu2021going} aim to simultaneously search and retrain the ensembles, which is beyond the affordable computational capacity for 3D medical images, where the input size is about 2.6k times larger than the CIFAR-10 dataset (\ie, $9\times96^3$ \textit{v.s.} $3\times32^2$). Third, most existing methods~\cite{zaidi2021neural,chen2021one} focus on the classification task, using the disagreement of base classifiers as the diversity indicator, which is not applicable for regression tasks such as dose prediction.
\section{Methods}
\label{sec:methods}
The framework of the proposed LENAS is shown in Fig.~\ref{fig_framwork}. 
It consists of two main components: 1) the U-NAS pipeline, a differentiable NAS for automatic architecture selection, and 2) KDA-Net, which transfers the knowledge from the U-NAS ensembles to a lightweight network through adversarial learning.
This enhances the models' ability to extract meaningful features. 
On the other hand, KDA-Net plays a crucial role in reducing inference time while maintaining competitive accuracy.
By combining these two components, U-NAS and KDA-Net, the LENAS framework achieves the dual objectives of enhancing feature extraction through diverse ensembles and optimizing inference efficiency without compromising accuracy. 
In the following sections, we provide a detailed introduction to each component.

\subsection{U-NAS}
As shown in Fig.~\ref{fig_framwork}, the proposed U-NAS follows the autoencoder~\cite{ronneberger2015u} structure with four down cells (DC) and four up cells (UC). 
Each individual cell is trained within an extensive search space comprising approximately 40,000 architecture configurations. 
In the following, we first introduce the search space and then describe the training strategy for joint optimization of the architecture and its weights.

\noindent\textbf{Search Space.}
The yellow and red blocks in Fig.~\ref{fig_framwork} show the network topologies of DC and UC, respectively, which include several fundamental computing units called hybrid modules (HMs). 
Each HM is a combination of different operations with four types: normal (N), downward (D), upward (U), and connect (C). These correspond to distinct operation groups within the search space.
As shown in Table~\ref{tab_ops}, we include the following operations in the search space: convolution (conv),  squeeze-and-excitation convolution (se\_conv), dilated convolution (dil\_conv), depthwise-separable convolution (dep\_conv), max pooling (max\_pool), average pooling (avg\_pool), trilinear interpolation (interpolate), and residual connection (identity). 

\begin{table}[htbp]
\caption{Operation set used for searching cells.}
\setlength{\tabcolsep}{4pt}{
\begin{tabular}{cccccc}
\toprule
NormOps   & DownOps         & UpOps           & {ConnectOps} & pre      & post                   \\ 
\midrule
identity  & avg\_pool       & up\_se\_conv    & {identity}   & conv  & conv  \\
conv      & max\_pool       & up\_dep\_conv   & {no connection}       &          &                        \\
se\_conv  & down\_se\_conv  & up\_conv        & {}           &          &                        \\
dil\_conv & down\_dil\_conv & up\_dil\_conv   & {}           &          &                        \\
dep\_conv & down\_dep\_conv & interpolate & &          &                        \\
          & down\_conv      &                 & &          &                        \\ 
\bottomrule
\end{tabular}}
\label{tab_ops}
\end{table}

The prefix `down' means the stride of the convolution operation is two, while the prefix `up' indicates the transposed convolution, which doubles the image resolution. 
For the first three columns of Table~\ref{tab_ops}, we use $3\times 3\times 3$ kernels for all convolution operations in the Conv-IN-ReLU order. 
In addition, a $3\times 3\times 3$ convolution (pre) and a $1 \times 1 \times 1$ convolution (post) are applied to adjust the number of channels.

\begin{algorithm}[thb]
\footnotesize
\renewcommand\arraystretch{1.2}
\caption{Training algorithm of LENAS}
    \begin{algorithmic}
    \Require Create the mixed operations $\overline{O}$ parameterized by $\alpha$; randomly initialized weights $\omega$; split the training set into two subsets: $\mathcal{D}_{\text{train}}$ and $\mathcal{D}_{\text{val}}$.
    \For{$m$ in $M$} \Comment{Search \& train $M$ teacher models}
    \While{not converged} \Comment{Search an architecture}
        \State 1. Update $\omega$ by gradient descending $\nabla _\omega \mathcal{L}_{\mathrm{dose}}(\omega , \alpha)$ on $\mathcal{D}_{\text{train}}$.
        \State 2. Update $\alpha$ by gradient descending $\nabla _{\alpha} \mathcal{L}_{\mathrm{nas}}(\omega , \alpha)$ on $\mathcal{D}_{\text{val}}$.
    \EndWhile
    \State Replace $\overline{O}$ with $O=O_i$, $i=\arg \max_k \exp (\alpha _k)/\sum ^{N}_{j=1} \exp(\alpha _j)$.
    \State \Comment{Derive the searched architecture.}
    \State Re-train the network with the best learned cell structures on $\mathcal{D}_{\mathrm{train}}$.
    \State $m \gets m+1$
    \EndFor \Comment{Finish the U-NAS stage}
    \State Train KDA-Net with $\mathcal{L}_{\text{KDA}}$. 
    \end{algorithmic}
    \label{algo_1}
\end{algorithm}

\noindent\textbf{Training Strategy.}
The training strategy of U-NAS contains two stages: the search process and the re-training process. 
In the search process, U-NAS is learned in a differentiable way \cite{liu2019darts}, which optimizes a super--network consisting of HMs with mixed operations. 
As Fig.~\ref{fig_framwork} shows, for each operation $O_i$ in total $N$ operations $O$, the weight of each operation is determined by the parameter $\alpha_i \in \alpha$, whose softmax transformation $\tilde{\alpha}_i=\exp(\alpha_i) / \sum^{N}_{j=1}\exp(\alpha_j)$ represents how much $O_i$ contributes to the HM. 
Then, the architecture parameters $\alpha$ and the network weights $\omega$ are learned by the mixed operations alternately. 
To explore different local optima, the search process is repeated multiple times with different initializations, leading to the discovery of diverse architectures.

Once the search process is completed, each HM retains only the most probable operation based on the parameter $\alpha$. 
Then, the DCs and UCs are replaced with the best-learned structures, and the network is re-trained on the training dataset $\mathcal{D}_{\mathrm{train}}$. 
The detailed training strategy of U-NAS is outlined in Algorithm~\ref{algo_1}.
During both the search and re-training processes, the difference between the predicted dose $\hat{y}$ and the target dose $y$ is measured using the $\mathcal{L}_1$ norm:
\begin{equation}
    \mathcal{L}_{\mathrm{dose}}=\| y - \hat{y} \|_1
    \label{eq_dose}
\end{equation}

\noindent\textbf{Diversity--Encouraging Loss.}
The diversity among the models obtained by U-NAS can be potentially achieved via different initializations. However, in many settings, the independent search process could converge to similar local optima as the same search goal is exploited. To optimize for diversity directly in the architecture search process, we propose a diversity--encouraging loss to encourage different predictions between the learned model with the best model.\footnote{The model with the best performance in multiple optimized architectures.}

In the search process of U-NAS, the primary objective is to achieve high accuracy for the learned model while also encouraging differences in predictions between the learned model and the best model. 
Consequently, during the training stage of U-NAS, the final loss function $\mathcal{L}{\mathrm{nas}}$ is formulated by combining the dose error loss $\mathcal{L}{\mathrm{dose}}$ (in Eq.~(\ref{eq_dose})) and the diversity--encouraging loss $\mathcal{L}_{\mathrm{div}}$ as follows:
\begin{equation}
    \begin{aligned}
    \mathcal{L}_{\mathrm{nas}} &=  \mathcal{L}_{\mathrm{dose}} +  \mathcal{L}_{\mathrm{div}} \\
     &= \|y - \hat{y}\|_1 + \eta \max (0, m - \frac{\left\|\hat{y}-\hat{y*} \right\|_1}{\left(\|\hat{y}\|_1 + \|\hat{y*}\|_1\right)/2} ),
    \end{aligned}
\end{equation}
where $\|\cdot\|_1$ is the voxel-wise $l_1$ norm; $y$, $\hat{y}$, and $\hat{y*}$ denote the ground-truth, prediction result of the training model and best model, respectively; $m$ is the margin (empirically set to 0.2) used to reduce the correlation between $\hat{y}$ and $\hat{y*}$ while avoiding the outliers; and $\eta$ is a weighting hyper-parameter to balance the two loss terms (empirically set to 1).

\subsection{KDA-Net}
The proposed KDA-Net performs knowledge distillation from the U-NAS ensembles to a single target network with adversarial learning. As shown in Fig.~\ref{fig_framwork}, we use a single U-Net network as the student and the average of multiple U-NAS predictions as the teacher ensemble. For all $K=8$ blocks (four D blocks and four C blocks) of the network, we apply the similarity loss on the intermediate output~\cite{zhang2020feature} between the teacher ensembles and the student based on the squared Euclidean distance:\footnote{Instead of ${L}_1$ loss in Eq.~\ref{eq_dose}, we adopt ${L}_2$ loss to the deep supervision for a fast optimization.}
\begin{equation}
    \mathcal{L}_{\mathrm{sim}}=\sum_{k=1}^8\left\|\frac{1}{M}\sum_{i=1}^M \left(I_k^{T_i} - I^S_k\right)\right\|_2^2,
    \label{eq_sim}
\end{equation}
where $I^{T_i}_k$ and $I^S_k$ denote the intermediate output of the $k$-th block of the $i$-th teacher network $T$ and student network $S$, respectively, and $M$ denotes the number of teacher networks.

To enhance the knowledge distillation process, we incorporate adversarial learning, which promotes the generation of similar features by both the student and the teachers. This is achieved by introducing a discriminator $D$ for each block. The discriminator is responsible for distinguishing between the outputs of the teachers and the student. By doing so, the student is encouraged to produce outputs that closely resemble those of the teachers.
The adversarial loss is defined as:
\begin{equation}
    \small\mathcal{L}_{\mathrm{adv}}=\sum_{k=1}^{8} \mathbb{E}_{I_k\sim P_T}\log D_k\left(I_k\right)+ \sum_{k=1}^{8} \mathbb{E}_{I_k\sim P_S}\log \left(1-D_k\left(I_k\right)\right),
    \label{eq_adv}
\end{equation}
where $I_k\sim P_T$ and $I_k\sim P_S$ denote the outputs from the $k$-th block of the teacher ensembles and the student network, respectively. Based on the above definition, we incorporate the dose loss in Eq.~(\ref{eq_dose}), the similarity loss in Eq.~(\ref{eq_sim}), and the adversarial loss in Eq.~(\ref{eq_adv}) into our KDA-Net loss function:
\begin{equation}
    \mathcal{L}_{\mathrm{KDA}}=\mathcal{L}_{\mathrm{dose}} + \lambda_1 \mathcal{L}_{\mathrm{sim}} + \lambda_2 \mathcal{L}_{\mathrm{adv}},
    \label{eq_final}
\end{equation}
where $\lambda_1$ and $\lambda_2$ are weighting hyper-parameters, which are empirically set to 0.05 and 0.01, respectively, in our experiments.
\section{Experiments}
\label{sec:experiments}
\subsection{Datasets}
\label{sec:dataset}
In this study, we evaluate the proposed method using two public datasets: the OpenKBP dataset and the AIMIS dataset.

\textbf{OpenKBP dataset.} The Open Knowledge-Based Planning (OpenKBP) dataset from the 2020 AAPM Grand Challenge~\cite{babier2020openkbp} is a public dataset consisting of 340 CT scans for the dose prediction task. The OpenKBP dataset includes scans of subjects being treated for head-and-neck cancer with radiation therapy. The data is partitioned into training ($n=200$), validation ($n=40$), and test ($n=100$) sets. The ROIs used in this study include the body, seven OARs (namely, brainstem, spinal cord, right parotid, left parotid, larynx, esophagus and mandible) and three planning target volumes (PTVs) with gross disease (PTV70), intermediate-risk target volumes (PTV63), and elective target volumes (PTV56).

\textbf{AIMIS dataset.} The AIMIS dataset from the 2021 Tencent AI Medical Innovation System (AIMIS) Challenge.\footnote{https://contest.taop.qq.com/channelDetail?id=108} Each scan is of a patient being treated for lung cancer with stereotactic body radiation therapy (SBRT). The dataset is officially partitioned into 300 scans for training, 100 scans for validation, and 100 scans for testing. The ROIs used in this study include the body and five OARs (namely, left lung, right lung, total lung, spinal cord, and heart), as well as an inner target volume (ITV) and planning target volume (PTV). 

\subsection{Implementation and Evaluation Metrics}
The pre-processing for the two datasets follows~\cite{liu2021cascade}. For normalization, the CT values are truncated to [-1024 HU, 1500 HU].
The following data augmentations are performed during training: horizontal and vertical flips, translation, and rotation around the $z$-axis.
For each sample of the OpenKBP dataset, the OAR masks (7 channels) and the merged target (1 channel) are concatenated with the CT scan (1 channel) as a $9\times 128\times 128\times 128$ tensor and fed into the dose prediction models.
For the AIMIS dataset, the input consists of OAR masks (5 channels), CT scan (1 channel), target volume (2 channels), and body (1 channel).

For the U-NAS search process, we first train the super--network for $8\times 10^4$ iterations using an Adam optimizer with an initial learning rate of $3\times 10^{-4}$, and a weight decay of $1\times 10^{-4}$. 
After that, the architecture parameters $\alpha$ are determined from the super network on the validation set. 
We repeat the search process multiple times with different random seeds to obtain various architectures. Then we re-train the searched models on the training set for $8\times 10^4$ iterations with a learning rate of $3\times 10^{-4}$.
For KDA-Net, we train the student network for $6\times 10^4$ iterations using an Adam optimizer with an initial learning rate of $1\times 10^{-5}$ and weight decay of $1\times 10^{-4}$. 

We use the official evaluation codes to validate the proposed method. 
Specifically, for the OpenKBP dataset, the evaluation metrics include (1) dose error, which calculates the mean absolute error (MAE) between the dose prediction and its corresponding ground-truth plan, and (2) DVH error, which calculates the absolute error of the DVH curves between the prediction and ground truth. According to~\cite{babier2020openkbp}, \{D99, D50 D1\} for the PTVs and \{D0.01cc, Dmean\} for the OARs are selected to measure the similarity of the DVH curves in this task.
For the AIMIS dataset, the evaluation is performed by measuring dose error with the mean squared error (MSE).
In addition, We use a paired t-test to calculate the statistical significance of the results.

\subsection{Experimental Results}
\subsubsection{Performance of U-NAS}
We first compare the performance of our U-NAS model with four manually designed architectures on the OpenKBP validation set. 
For each manually designed architecture, we employ the same convolution operation choice in HM, including \textit{conv}, \textit{se\_conv}, \textit{dil\_conv}, and \textit{dep\_conv}. 
We apply \textit{max\_pool}, \textit{interpolate}, and \textit{no connection} operations for all the manually designed architectures. 
Table~\ref{tab_NAS} shows a performance comparison of different models. Our U-NAS model outperforms all the manually designed networks on the body, seven OARs, and three PTVs. 
The single U-NAS model achieves an MAE of 2.580 and 1.736 in dose score and DVH score, respectively, outperforming the best manually designed network by 0.111 and 0.128 in dose error and DVH error, respectively.
It is interesting, in most cases, the ensemble of four models outperforms the corresponding individual models (for both the manually designed and NAS learned models), and the ensemble of the NAS models outperforms the ensemble of the manually designed models. Please refer to Sec.~\ref{sec:dis} for more discussions of ensemble learning.

\begin{table*}[thb]
\renewcommand\arraystretch{1.2}
\caption{Performance comparison of NAS models with manually designed networks on $\mathcal{D}_{val}$. BS, SC, RP, LA, ES, LE, MD, P$_{70}$, P$_{63}$, and P$_{56}$ denote Brainstem, Spinal Cord, Right Parotid, Left Parotid, Larynx, Esophagus, Mandible, PTV70, PTV63, and PTV56, respectivily. $\dagger$ represents significantly different results ($p < 0.05$, paired t-tests)}
\resizebox{1\linewidth}{!}{
\begin{tabular}{c|c|ccccc|cc}
\hline
\hline
\multicolumn{2}{c|}{\multirow{3}{*}{}}       & \multicolumn{5}{c|}{Single Model} & \multicolumn{2}{c}{Ensemble}\\ 
\cline{3-9} 
\multicolumn{2}{c|}{}                        & \multicolumn{1}{c}{Conv} & \multicolumn{1}{c}{Se\_conv} & \multicolumn{1}{c}{Dil\_conv} & \multicolumn{1}{c}{Dep\_conv} & NAS             & \multicolumn{1}{c}{Manual}                           &    NAS                                 \\ 
\hline 
\multicolumn{2}{c|}{Flops (G)}    & 160.52   & 57.33   & 160.52   & 63.52   & 111.68   & 441.89 & 398.59\\
\multicolumn{2}{c|}{Params (M)}    & 17.68   & 8.29   & 17.68   & 8.69   & 13.09   & 52.34  & 47.96\\
\cdashline{1-9}[0.8pt/2pt]
\multirow{11}{*}{\rotatebox{90}{Dose Error}} & Body & $2.634\pm 0.760$  & $2.736\pm 0.861\dagger$  & $2.741\pm 0.805\dagger$  & $2.728\pm 0.789\dagger$  & \bm{$2.581\pm 0.784$}  & $2.503\pm 0.762\dagger$  & \bm{$2.400\pm 0.743$} \\
                             & BS  & $1.606\pm 1.076$  & $1.646\pm 1.152$  & $1.743\pm 1.305\dagger$  & $1.677\pm 1.278$  & \bm{$1.486\pm 0.971$}  & $1.541\pm 1.136$  & \bm{$1.442\pm 1.035$} \\
                             & SC  & $2.095\pm 1.014$  & $2.131\pm 1.181$  & $2.184\pm 1.115$  & $2.068\pm 0.986\dagger$  & \bm{$2.018\pm 0.936$}  & $1.939\pm 0.909$  & \bm{$1.891\pm 0.878$} \\
                             & RP  & \bm{$3.040\pm 0.870$}  & $3.277\pm 1.020$  & $3.152\pm 0.921$  & $3.472\pm 1.117$  & $3.074\pm 0.909$  & $2.942\pm 0.868$  & \bm{$2.820\pm 0.889$} \\
                             & LA  & $3.154\pm 0.839$  & $3.208\pm 0.979$  & $3.075\pm 0.803$  & $3.383\pm 1.212$  & \bm{$3.029\pm 0.874$}  & $2.917\pm 0.740$  & \bm{$2.710\pm 0.753$} \\
                             & ES  & \bm{$2.428\pm 1.004$}  & $2.773\pm 0.830$  & $2.749\pm 1.036$  & $2.531\pm 1.077$  & $2.467\pm 1.009$  & $2.410\pm 0.892$  & \bm{$2.182\pm 0.705$} \\
                             & LE  & $3.034\pm 1.385$  & $3.204\pm 1.419$  & $3.451\pm 1.561$  & $3.148\pm 1.481$  & \bm{$2.883\pm 1.366$}  & $2.973\pm 1.334$  & \bm{$2.593\pm 0.727$} \\
                             & MD  & $3.988\pm 1.341$  & $3.992\pm 1.413$  & $4.086\pm 1.146$  & $4.051\pm 1.218$  & \bm{$3.800\pm 1.189$}  & $3.745\pm 1.152$  & \bm{$3.520\pm 1.133$} \\
                             & P$_{70}$  & $2.062\pm 1.004\dagger$  & $2.090\pm 1.167\dagger$  & $2.198\pm 1.463\dagger$  & $2.045\pm 0.895\dagger$  & \bm{$1.620\pm 0.789$}  & $1.896\pm 1.108\dagger$  & \bm{$1.570\pm 0.784$} \\
                             & P$_{63}$  & $2.345\pm 1.137$  & $2.534\pm 1.250\dagger$  & $2.686\pm 1.566\dagger$  & $2.521\pm 1.090\dagger$  & \bm{$2.135\pm 0.950$}  & $2.318\pm 1.213\dagger$  & \bm{$2.057\pm 0.926$} \\
                             & P$_{56}$ & $2.227\pm 0.916$  & $2.296\pm 0.885$  & $2.326\pm 1.029\dagger$  & $2.371\pm 0.758\dagger$  & \bm{$2.116\pm 0.768$}  & $2.122\pm 0.841\dagger$  & \bm{$1.926\pm 0.703$} \\ 
\hline
\hline
\end{tabular}}
\label{tab_NAS}
\end{table*}

\begin{figure}[thbp]
    \centering
    \subfloat[]{
		\includegraphics[width=0.98\linewidth]{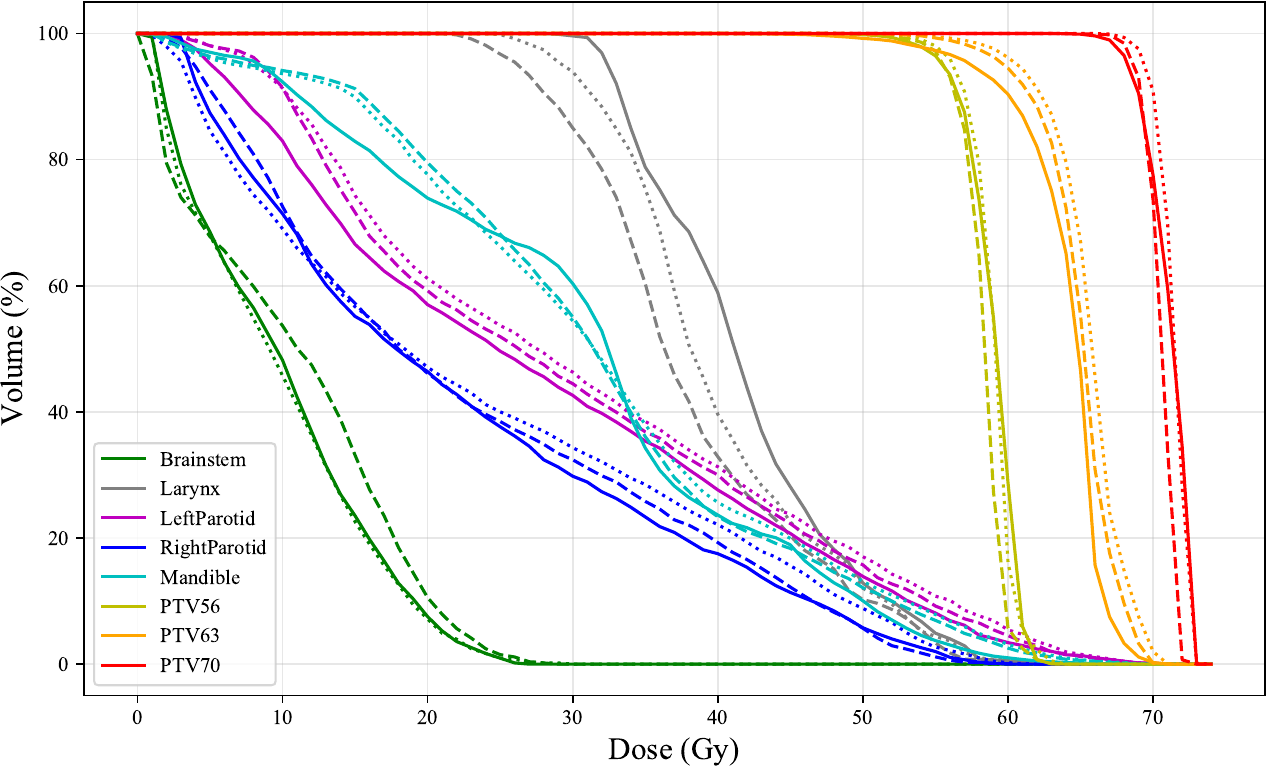}
		\label{fig:multi_dvh}
	} \\
    \subfloat[]{
		\includegraphics[width=\linewidth]{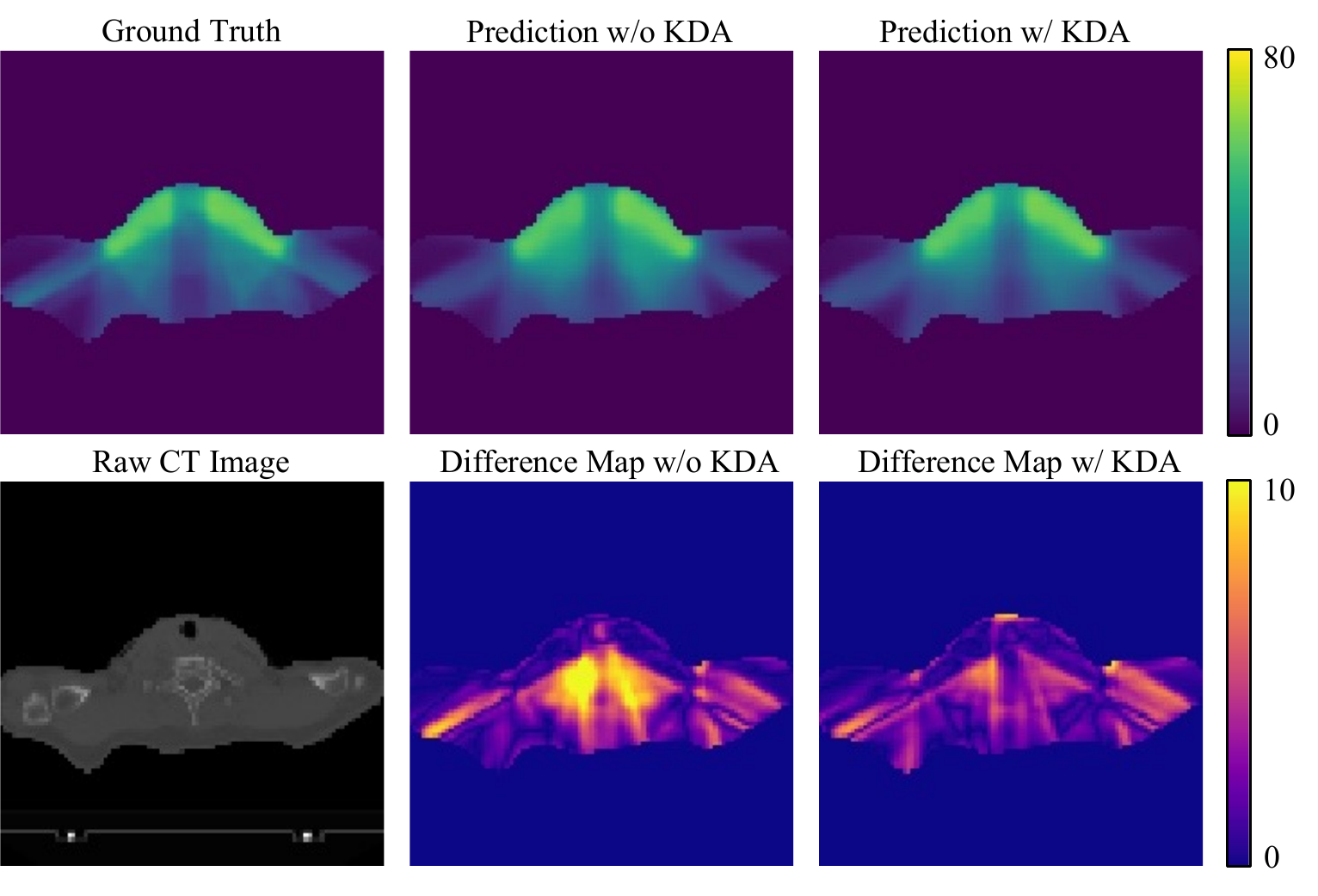}
		\label{fig:visual_kda}
	}
    \caption{(a) DVHs of the dose distribution of ground-truth plan (solid curves) and predictions by a single U-Net with and without the proposed KDA method, illustrated by dashed and dotted lines, respectively. The PTV70, PTV63, and PTV56 are shown in red, orange, and yellow lines, respectively. (b) An example of dose distributions of the clinical plan and predicted plans of the single U-Net with and without the KDA method.}
    \label{fig:kda}
\end{figure}

\subsubsection{Performance of KDA-Net}
We compare the performance of a single U-Net with and without the proposed KDA module, including the dose distributions and DVH, on the OpenKBP validation set.
Fig.~\ref{fig:multi_dvh} shows an example of the DVH curves from a patient of the validation set. 
The solid lines represent the DVH curves of the ground truth, while the dashed and dotted lines represent the DVHs extracted from the predicted dose of U-Net with and without KDA (\ie, train from scratch), respectively.
For this example patient, U-Net with KDA exhibits a better agreement in predicting the dose to the PTVs. The predictions of OARs are more variable between the two methods.
Fig.~\ref{fig:visual_kda} shows the corresponding dose color contour for the same patient as in Fig.~\ref{fig:multi_dvh}, which suggests that the single U-Net model with KDA is able to achieve better dosimetric congruence with the original plan on the PTV.

\begin{table}[!t]
\caption{Comparison of performance with the state-of-the-art methods on the OpenKBP test set.}
\centering
\renewcommand\arraystretch{1.2}
\setlength{\tabcolsep}{9pt}{
\begin{tabular}{l|l|ll|ll}
\hline
\hline
\multicolumn{2}{c|}{\multirow{2}{*}{Methods}}         & \multicolumn{2}{c|}{Dose Score} & \multicolumn{2}{c}{DVH Score} \\
\cline{3-6}
\multicolumn{2}{c|}{}           & MAE    & MSE    & MAE   & MSE\\
\hline
\multirow{5}{*}{\rotatebox{90}{Leaderboard}
}    &Top \#1    & 2.429 & 15.488 & 1.478 & 5.913\\
                                &Top \#2    & 2.564 & 16.550 & 1.704 & 6.812\\
                                &Top \#3    & 2.615 & 17.774 & 1.582 & 6.961\\
                                &Top \#4    & 2.650 & 18.091 & 1.539 & 6.031\\
                                &Top \#5    & 2.679 & 18.023 & 1.573 & 6.525\\
\cdashline{1-6}[0.8pt/2pt]
\multirow{6}{*}{\rotatebox{90}{Single Model}}   &FCN~\cite{long2015fully}  & 2.681 & 18.144 & 2.452 & 12.310 \\
                                &V-Net~\cite{milletari2016v} & 3.129 & 23.336 & 2.325 & 11.417 \\
                                &U-Net~\cite{ronneberger2015u} & 2.619 & 17.221 & 2.313 & 11.343 \\
                                &ResUNet~\cite{yu2017volumetric} & 2.601   & 16.932   & 2.209   & 10.591 \\
                                &U-NAS (ours) & 2.597 & 16.962 & 1.803 & 7.628 \\
                                &LENAS (ours)  & \textbf{2.565} & \textbf{16.614} & \textbf{1.737} & \textbf{7.272} \\
\cdashline{1-6}[0.8pt/2pt]
\multirow{3}{*}{\rotatebox{90}{Cascade}}        &U-Net~\cite{ronneberger2015u} & 2.461   & 15.489    & 1.588   & 6.511 \\
                                &ResUNet~\cite{yu2017volumetric}   & 2.448   & 16.023    & 1.499   & 5.855 \\
                                &U-NAS (ours) & \textbf{2.434}   & \textbf{15.376}     & \textbf{1.496}   & \textbf{5.564} \\
\cdashline{1-6}[0.8pt/2pt]
\multirow{2}{*}{\rotatebox{90}{Ens.}}       &Off-the-shelf & 2.521   & 16.060    & 1.771   & 6.851 \\
                                &U-NAS (ours) & \textbf{2.357}   & \textbf{14.326}    & \textbf{1.465}   & \textbf{5.560} \\
\hline
\hline
\end{tabular}}
\label{tab_sota}
\end{table}

\begin{table}[!t]
\renewcommand\arraystretch{1.2}
    \centering
    \caption{Comparison with the state-of-the-art methods on the test set of AIMIS.
    }
    \setlength{\tabcolsep}{3pt}{
    \begin{tabular}{c|ll|c|ll}
        \hline
        \hline
        \multicolumn{3}{c|}{Primary Phase} & \multicolumn{3}{c}{Final Phase} \\
        \hline
        Rank & Team & Dose Score & Rank & Team & Dose Score\\
        \hline
        \textbf{\#1}  & \textbf{qqll (ours)}      & \textbf{15611.6398} & \textbf{\#1}  & \textbf{qqll (ours)}   & \textbf{15571.6051} \\
        \#2  & deepmedimg     & 17223.3940 & \#2  & gosnail      & 15869.4256 \\
        \#3  & gosnail         & 18425.5708 & \#3  & teamC        & 16323.9720 \\
        \#4  & adosepredictor & 18638.4767 & \#4  & 27149         & 16486.1417 \\
        \#5  & star             & 19340.0643 & \#5  & capsicummeat & 18137.9836 \\
        \hline
        \hline
    \end{tabular}}
    \label{tab:aimis_test}
\end{table}

\begin{table}[!t]
\renewcommand\arraystretch{1.2}
    \centering
    \caption{Comparison of U-NAS with the off-the-shelf models on the validation set of AIMIS.}
    \footnotesize
    \resizebox{\linewidth}{!}{
    \begin{tabular}{c|ccccccccc}
    \hline
    \hline
        \multirow{2}{*}{Methods} & \multirow{2}{*}{All} & \multirow{2}{*}{Body} & \multirow{2}{*}{Heart} & \multirow{2}{*}{L-Lung} & \multirow{2}{*}{R-Lung} & Total & Spinal & \multirow{2}{*}{ITV} & \multirow{2}{*}{PTV} \\
        & & & & & & -Lung & -Cord & &  \\
    \hline
        U-Net    & 9801 & 56608 & 45643 & \textbf{71894} & 75099 & \textbf{64108} & \textbf{68377} & 525499 & \textbf{842770} \\
        ResUNet & 9782 & 56668 & \textbf{41858} & 77288 & 77399 & 66066 & 71790 & 593486 & 904382 \\
        U-NAS    & \textbf{9484} & \textbf{54839} & 43746 & 82291 & \textbf{71597} & 66922 & 71175 & \textbf{510381} & 858750 \\
    \hline
    \hline
    \end{tabular}}
    \label{tab:aimis_val}
\end{table}

\subsubsection{Comparison with the State-of-the-art Methods}
In Table~\ref{tab_sota}, we compare the proposed LENAS model with several state-of-the-art methods on the OpenKBP test set. 
The competing methods include 3D FCN~\cite{long2015fully}, V-Net~\cite{long2015fully}, 3D U-Net~\cite{ronneberger2015u}, 3D ResUNet~\cite{yu2017volumetric}, and five top-ranking methods on the AAPM-2020 challenge leaderboard~\cite{babier2020openkbp}.
We thoroughly compare our LENAS model with existing methods using single model, cascade, and ensemble strategies.
The cascade strategy is to sequentially combine two networks and produce the results in a coarse-to-fine fashion.
\textbf{For single model}, our U-NAS achieves an MAE of 2.597 and 1.803 in dose score and DVH score, respectively, outperforming the best off-the-shelf method (ResUNet). Integrating the KDA (\ie, LENAS) further improve the performance to 2.565 and 1.737, respectively. 
\textbf{For cascade models}, our cascade U-NAS model achieves 2.434 and 1.496 MAE in dose score and DVH score, respectively, outperforming the cascade ResUNet which achieves 2.448 and 1.499.
\textbf{For five model ensembles},
our U-NAS ensemble achieves 2.357 MAE and 14.326 MSE in dose score, and 1.465 MAE and 5.560 MSE in DVH score, outperforming the ensembles of the off-the-shelf models and the top-ranking solutions on the AAPM-2020 challenge leaderboard.

\subsubsection{Generalization Evaluation}
We further explore the generalizabilty of our method on the AIMIS dataset. Specifically, we apply the best architecture (single model) learned from the OpenKBP dataset to the AIMIS challenge. The evaluation results are calculated by the organizers\footnote{https://contest.taop.qq.com} of the challenge, and shown in Table~\ref{tab:aimis_test}.
Our U-NAS method achieves first place in the AIMIS challenge in both the primary and final phases,\footnote{In the primary phase, the test set consists of 50 scans, and in the final phase, the test set consists of another 150 scans.} outperforming the runner-up by 9.36\% and 1.88\%, respectively. 
Moreover, in Table~\ref{tab:aimis_val}, we further compare our U-NAS model with two best performing off-the-shelf models, U-Net and ResUNet, with respect to different ROIs.
A consistent trend can be observed that our U-NAS outperforms the off-the-shelf models and other top-ranking solutions on both the validation set and the test set of AIMIS.

\begin{figure}[thb]
    \centering
    \subfloat[]{
		\includegraphics[width=0.35\textwidth]{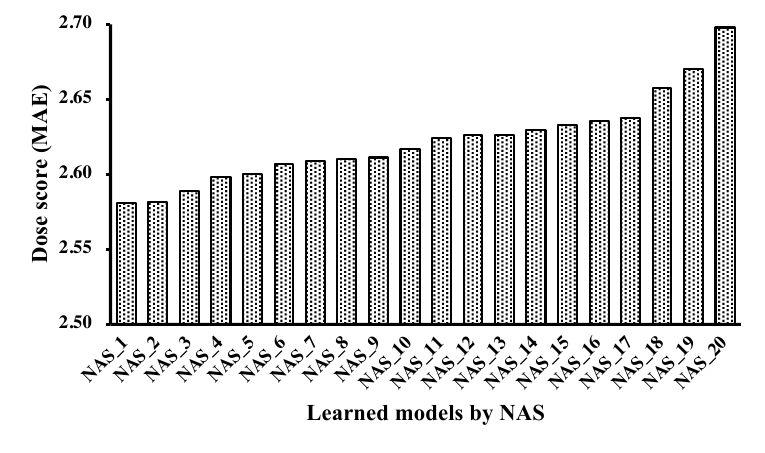}
		\label{fig:mba_nas_20}
	}\\
    \subfloat[]{
		\includegraphics[width=0.35\textwidth]{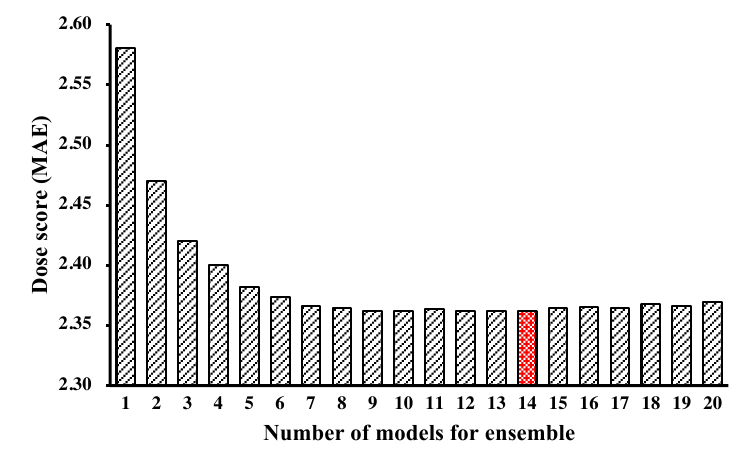}
		\label{fig:mba_nas_20_en}
	}
    \caption{The dose score (MAE) of (a) 20 NAS models; (b) the ensembles.}
    \label{fig:many_better_all}
\end{figure}
\section{Discussions}
\label{sec:dis}
In this section, we investigate the correlations between diversity and ensemble performance, and empirically provide insightful guidance for ensemble learning with NAS in the task of dose prediction. 

\subsection{Ensemble Many is Better than All}
In ensemble learning, most methods brusquely integrate all the obtained models to obtain the final result.
To explore the correlation between the number of ensembles and their corresponding performance, we follow~\cite{zhou2002ensembling} to systemically conduct the search process multiple times and select the top 20 models for the experiment. 
Then, we average the results one by one sequentially w.r.t. the individual performance. The results are shown in Fig.~\ref{fig:many_better_all}.
Fig.~\ref{fig:mba_nas_20} shows the dose score of the 20 selected models, which ranges from 2.5806 (NAS\_1) to 2.6979 (NAS\_20) MAE.
Fig.~\ref{fig:mba_nas_20_en} shows the dose score of ensembles of the top--$k\in[1,20]$ models. It can be seen that the ensembles achieve the best performance with the top 14 models (2.3621 MAE) instead of all the 20 models (2.3693 MAE).
Intuitively, the inclusion of models with unacceptable performance in the ensembles could hurt the final ensemble results.
Thus, next step is to explore the selection criteria for the members of the ensemble.

\begin{figure}[!t]
    \centering
    \includegraphics[width=0.45\textwidth]{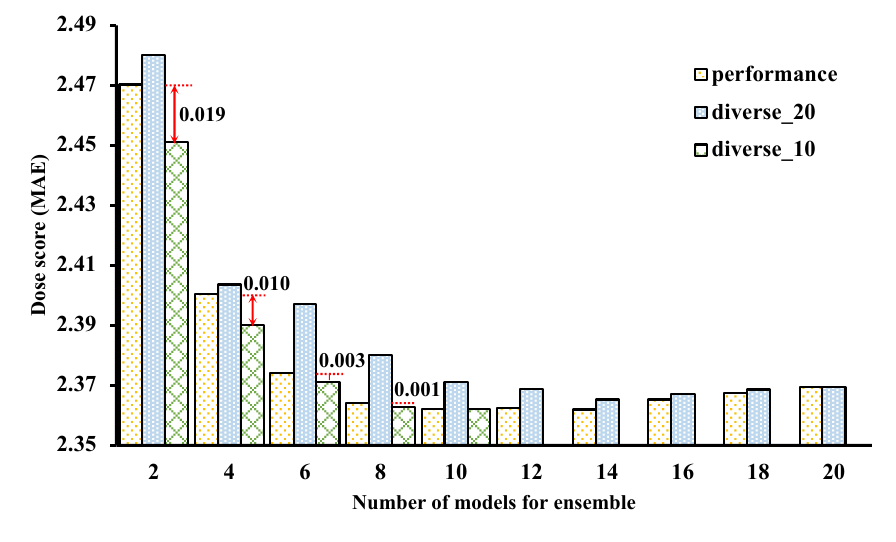}
    \caption{The dose score (MAE) of the ensembles with different numbers of models. The yellow bars indicate the models selected based on performance; the blue and green bars indicate the models selected based on diversity from the top 20 models and ten models, respectively.}
    \label{fig:per_div}
\end{figure}

\subsection{Performance vs. Diversity}
Extensive literature~\cite{bian2021does} has shown that the ideal ensemble is one comprised of accurate networks that make errors on diﬀerent parts of the input space. 
In other words, the performance of an ensemble depends not only on the accuracy of base learners but also on their diversity.
However, typically, existing methods implicitly encourage the diversity of ensembles with different training strategies (\eg, random initializations, different hyper-parameters and loss functions), then causally select the models based on the accuracy. 
So, \textit{does diversity matter when selecting the members of the ensemble?}
To answer the question, we conduct the experiments.

The diversity between two models is quantified by calculating the average root mean absolute error (RMAE) between their predictions: 
$d(y_a, y_b) = \frac{|y_a^i - y_b^i|}{y_a^i + y_b^i}$,
where $y_a$ and $y_b$ are the outputs of two models. 
Based on the performance and diversity of individual models, we select different pairs of models for ensembling. 
The yellow and blue bars in Fig.~\ref{fig:per_div} illustrate that, for the total 20 models, ensembling based on the performance of individual models consistently outperforms ensembling based on diversity alone. 
It reveals that \textit{the performance of individual model's performance is an essential factor in ensembling.}
In addition, the yellow and green bars show that, for the top 10 models, the results exactly opposite, as the ensemble performances based on the individual models' performance is lower than that based on the diversity.
This suggests that diversity is indeed an important factor as well. In particular, \textit{when the performances of the individual models are comparable, the diversity is more important than the accuracy.}

\begin{figure}[!t]
    \centering
    \subfloat[]{
		\includegraphics[width=0.49\linewidth]{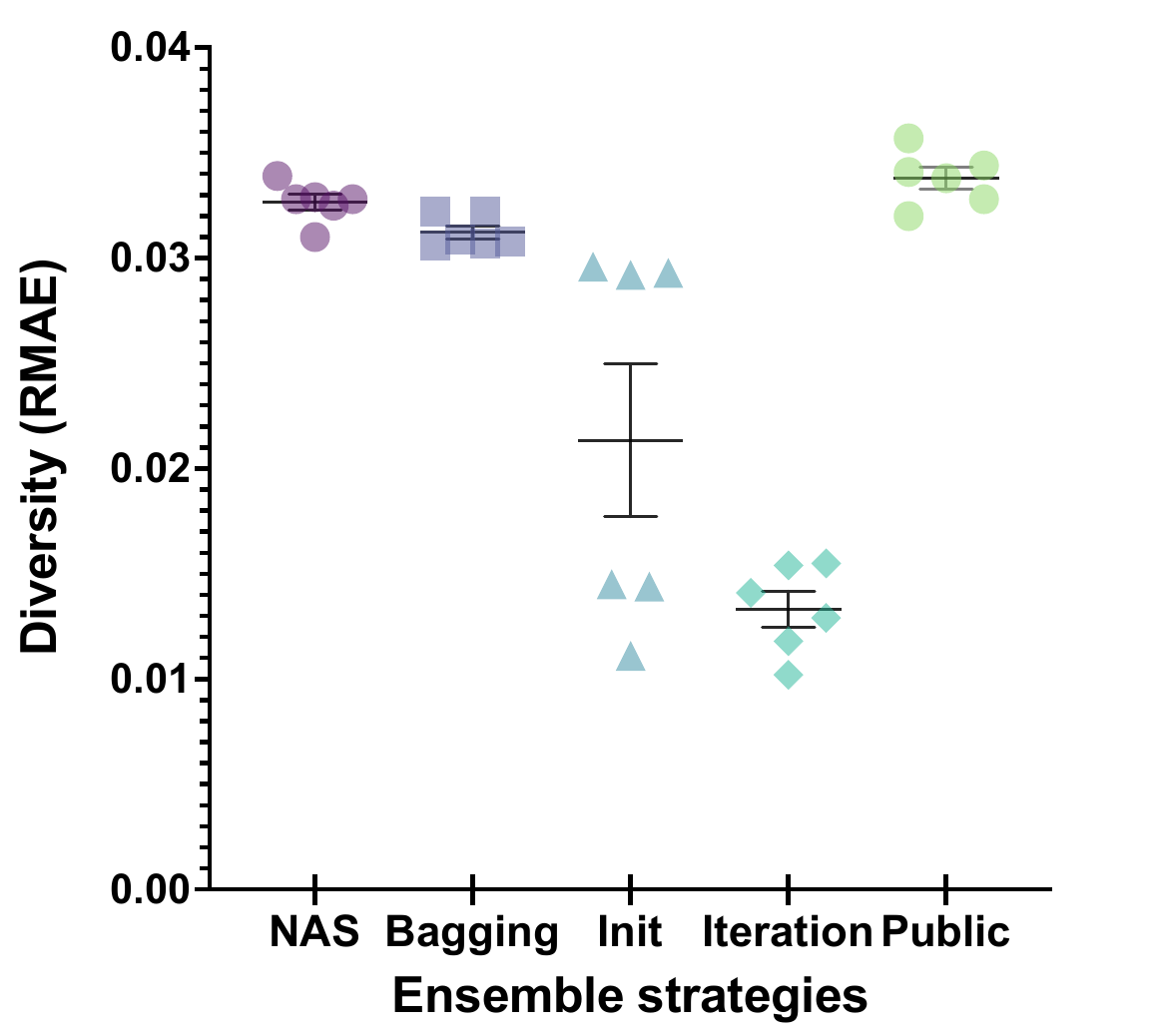}
		\label{fig:strategies_div}
	}
    \subfloat[]{
		\includegraphics[width=0.49\linewidth]{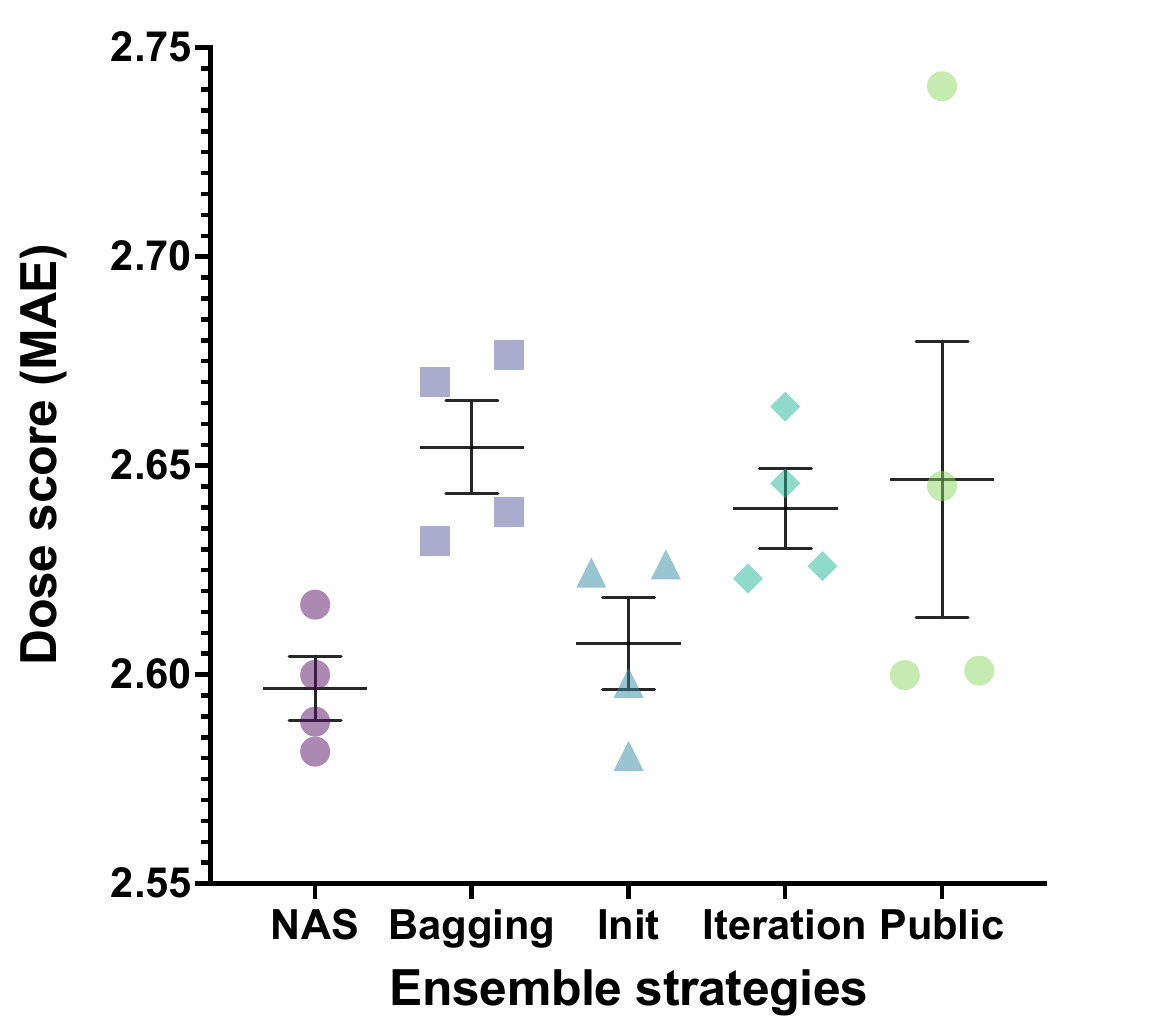}
		\label{fig:strategies_dose}
	}
    \caption{(a) Diversity and (b) dose score (MAE) of individual models in different ensemble strategies: NAS, bagging, random initializations and iterations, and different off-the-shelf architectures (denotes public).}
    \label{fig:diff_strate}
\end{figure}

\begin{figure}[!t]
    \centering
    \includegraphics[width=0.35\textwidth]{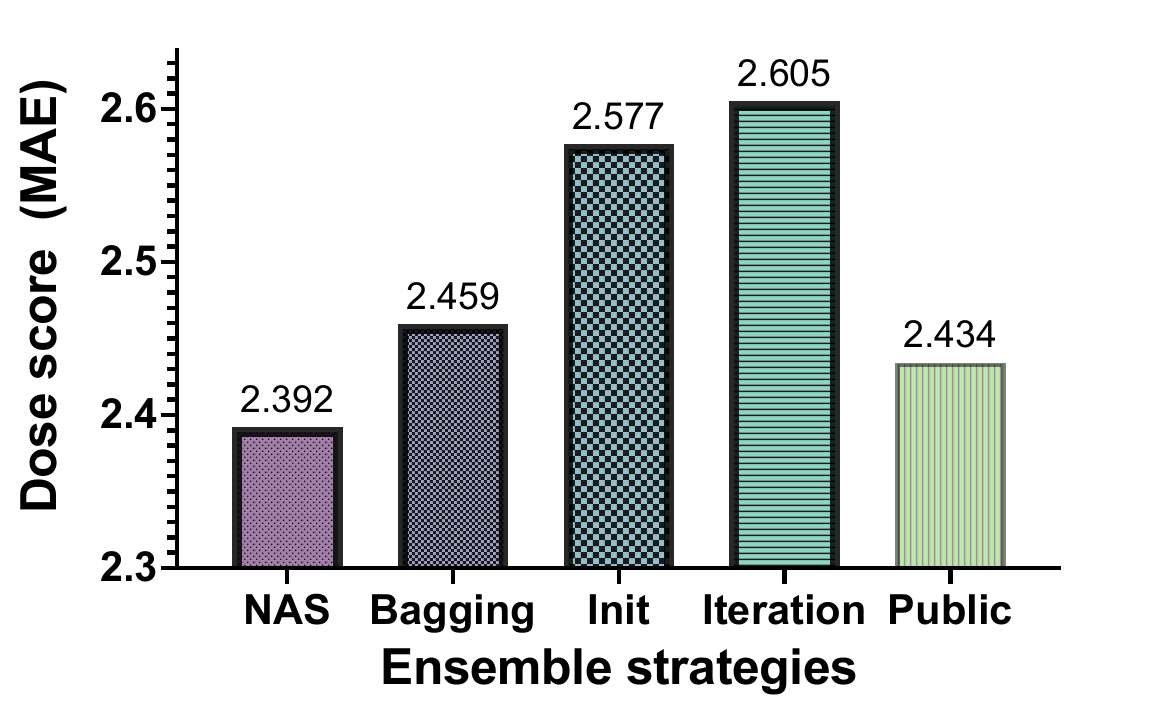}
    \caption{Overall Dose score (MAE) of the ensembles with different strategies.}
    \label{fig:diff_strate_en}
\end{figure}

\subsection{Comparison of Different Ensemble Strategies}
The uniqueness of the proposed LENAS model is to exploit a diverse collection of network structures that drive a natural bias towards the diversity of predictions produced by individual networks. 
To assess the impact of LENAS on the ensembling, we compare the diversity and performance of our method with four ensembling strategies, including bagging, random initializations, different training iterations, and different off-the-shelf architectures, on the OpenKBP validation set, as shown in Fig.~\ref{fig:diff_strate}.
For each strategy, we obtain four models randomly. 
Specifically, for the NAS models, we select the top four models from the aforementioned 20 models (NAS\_1 to NAS\_4). 
For bagging, we split the training set into four non-overlapped subsets using different portions of the data (three subsets) to train the four models.
For random initialization, we repeat the training procedures four times with different initialization seeds. For the different training epochs, we train a single network with $8 \times 10^4$ iterations, and pick the last four checkpoints with a gap of 2000 training iterations in between. 
Finally, for the off-the-shelf architectures, we select the four most popular architectures in 3D medical image analysis, including FCN, VNet, U-Net, and ResUNet.

The diversities of the four ensembling strategies are illustrated in Fig.~\ref{fig:strategies_div}. 
The diversity of the NAS models is 0.0326 RMAE with a standard deviation of 0.0009, greater than three of the other strategies, namely cross-validation, random initialization and iterations, and comparable to off-the-shelf architectures, which achieve a diversity of $0.0338\pm0.0013$.
Furthermore, the results presented in Fig.~\ref{fig:strategies_dose} demonstrate that the NAS models outperform other strategies by a significant margin in terms of the mean and standard deviation of the dose score, with a value of $2.587\pm0.0082$.
Although the diversity of 4-fold cross-validation is close to the NAS models, the individual models' performance suffers from the limited representations of the subsets of training data, leading to a lower ensemble performance.
Similar trends are observed in off-the-shelf models.
Lastly, the performance of the ensembles, as measured by the dose score in terms of MAE, is depicted in Fig.~\ref{fig:diff_strate_en}. 
It is evident that the ensembles of NAS models achieve the best performance, with an MAE of 2.392, surpassing other ensemble strategies. 
This result emphasizes the superiority of producing and ensembling different network architectures as opposed to creating an ensemble consisting solely of duplicates of a single network architecture with different model parameters.

\begin{figure}[!t]
    \centering
    \subfloat[]{
		\includegraphics[width=0.48\textwidth]{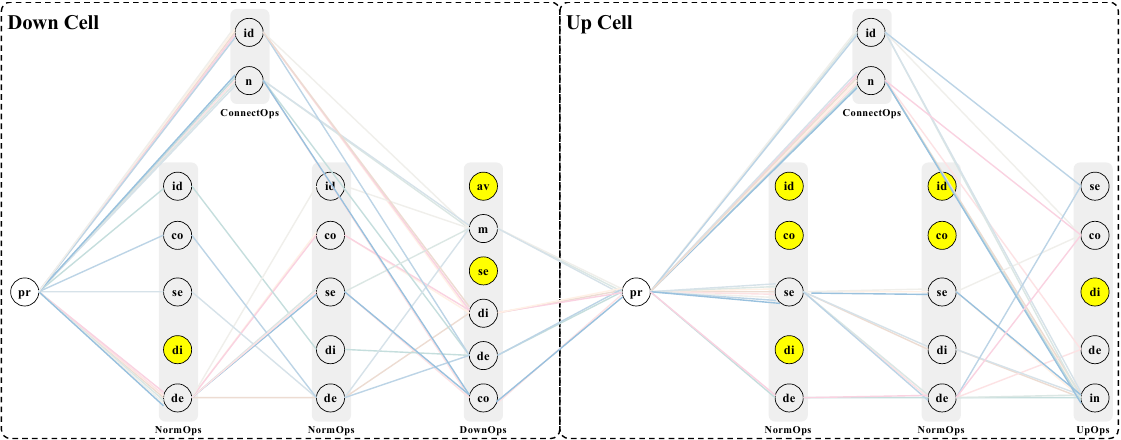}
		\label{fig:wo_loss}
	}\\
    \subfloat[]{
		\includegraphics[width=0.48\textwidth]{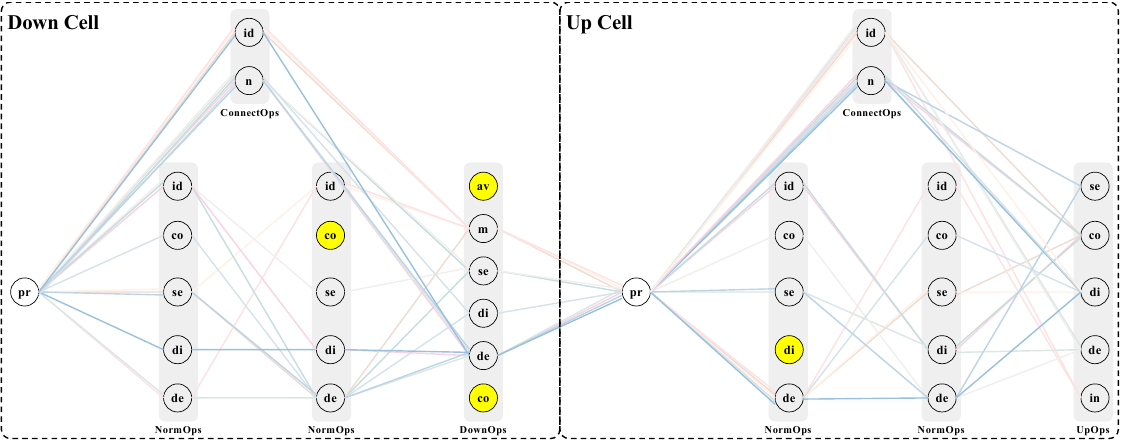}
		\label{fig:w_loss}
	}
    \caption{The DCs and UCs of ten learned architectures (a) without and (b) with diversity--encouraging loss, where \textit{id}, \textit{n}, \textit{co}, \textit{se}, \textit{di}, \textit{de}, \textit{av}, \textit{m}, \textit{in}, and \textit{pr} denote \textit{identity}, \textit{no connection}, \textit{conv}, \textit{se\_conv}, \textit{dil\_conv}, \textit{dep\_conv}, \textit{avg\_pool}, \textit{max\_pool}, \textit{interpolate}, and \textit{pre} in Table~\ref{tab_ops}, respectively. The yellow operations are not used by the ten architectures.}
    \label{fig:div_loss}
\end{figure}

\subsection{Effectiveness of Diversity--Encouraging Loss}
We investigate the effectiveness of the diversity--encouraging loss in Fig.~\ref{fig:div_loss}. 
Specifically, we search ten architectures with and without diversity--encouraging loss,\footnote{We follow~\cite{liu2019darts} to search single architecture of DC and UC in each model for facilitating the optimization of the search process.} and the learned DCs and UCs are shown in Fig.~\ref{fig:div_loss}. We further calculate the variation of the operations in each module. 
Specifically, we first rank the operations in each HM based on their frequency in all the architectures (\eg, the operation with the highest frequency is identified with ID 0), and then produce all the standard deviations of each HM.
The quantitative results of the variation with and without diversity--encouraging loss are 328.3 and 31.1, respectively, indicating that with the diversity--encouraging loss, the U-NAS method can generate architectures with a greater variation, and consequently encourage the diversity of the predictions.
\section{Conclusion and Future Work}
In this paper, we proposed a learning-based ensemble approach, named LENAS, for 3D radiotherapy dose prediction. 
The two key components of LENAS are 1) a U-NAS framework which automatically searches neural architectures from numerous architecture configurations to form a teacher network zoo, and 2) a KDA-Net which hierarchically transfers the knowledge distilled from the teacher networks to the student network to reduce the inference time while maintaining competitive accuracy. 
We conducted comprehensive experiments to investigate the impact of diversity in ensemble learning, and derived several empirical guidelines for producing and ensembling base learners in consideration of individual accuracy and diversity.
Extensive experiments on two public datasets demonstrated the effectiveness and superior performance of our method compared to state-of-the-arts. 

The LENAS method offers several notable advantages.
Firstly, the U-NAS framework enables automatic neural architecture search, allowing for the efficient exploration of a vast array of architecture configurations. This search process results in the creation of a diverse teacher network zoo, which in turn facilitates the generation of a robust ensemble of learners. By leveraging this ensemble method, we are able to enhance the overall performance of the system.
Secondly, the KDA-Net enables a hierarchical transfer of distilled knowledge from the teacher networks to the student network. This transfer mechanism not only reduces inference time but also ensures that the student network maintains a high level of accuracy, thus striking an optimal balance between efficiency and performance.
Lastly, the effectiveness of our approach has been thoroughly validated through rigorous experimentation on two public datasets. Notably, our method has achieved first place in both the AAPM and AIMIS challenges. This outstanding performance serves as compelling evidence of the promising potential and efficacy of our proposed LENAS method.

We would like to point out several limitations of our work. 
First, the NAS ensembles require multiple rounds of searching-retraining, which is very time-consuming in the training phase. 
Second, a few failure models may be generated by NAS. 
This situation is also common in the gradient-based NAS methods. 
Third, the diversity between learners in an ensemble is hard to formulate appropriately, which could be task-specific and vary for different outputs (\eg, classification, segmentation, and regression).
Future studies will be focused on: 
1) a more specific model selection criterion for the best ensemble strategies; 
2) a computational-efficient training strategy for multiple ensemble learners;
and 3) an optimization method from dose prediction map to the radiotherapy treatment plan.
\bibliographystyle{IEEEtran}
\bibliography{refs}

\end{document}